\documentclass[aps,pre,showpacs,groupedaddress]{revtex4}

\usepackage{graphicx,psfrag}

\def\BibTeX{{\rm B\kern-.05em{\sc i\kern-.025em b}\kern-.08em
    T\kern-.1667em\lower.7ex\hbox{E}\kern-.125emX}}

\begin{document}

\title{Parametric characterisation of a chaotic attractor using two scale Cantor measure}

\author{K. P. Harikrishnan}
\email{kp_hk2002@yahoo.co.in}
\affiliation{Department of Physics, The Cochin College, Cochin-682 002, India} 
\author{R. Misra}
\affiliation{Inter University Centre for Astronomy and Astrophysics, Pune-411 007, India} 
\author{G. Ambika}
\email{g.ambika@iiserpune.ac.in}
\affiliation{Indian Institute of Science Education and Research, Pune-411 021, India} 
\author{R. E. Amritkar}
\affiliation{Physical Research Laboratory, Navarangapura, Ahmedabad-380 009, India} 

\begin{abstract}
A chaotic attractor is usually characterised by its multifractal spectrum which  gives a  
geometric measure of its complexity. Here we present a characterisation using a minimal  set of 
independant parameters which are uniquely determined by  the underlying process that 
generates the attractor. The method maps the $f(\alpha)$ spectrum of a 
chaotic attractor onto that of a general two scale Cantor measure. We show that the mapping can 
be done  for a large number of chaotic systems. In order to implement this procedure, 
we also 
propose a generalisation of the standard equations for two scale Cantor set in one dimension 
to that in higher 
dimensions. Another interesting result we have obtained both theoretically and numerically 
is that, 
the $f(\alpha)$ characterisation gives information  only upto two scales,   
even when the underlying process generating the multifractal involves more than two scales.
\end{abstract}


\maketitle

\section{\label{sec:level1}INTRODUCTION}
The existence of a multifractal measure for any system most often indicates an underlying 
process generating it, 
be it multiplicative or dynamic. In the context of chaotic attractors  arising from 
dynamical systems, their multifractal measure result from a time ordered process, 
which may be an iterative scheme or a continuous flow \cite {eck}. The description of 
the  invariant measures in terms of $D_q$ \cite {hen} or $f(\alpha)$ \cite {hal3},  
however, provides only a 
characterisation of their geometric complexity. Feigenbaum et.al \cite {fei,feig} 
and Amritkar and Gupte \cite {gup} have 
shown that it is also possible to get the dynamical information in some specific 
cases by inverting the information contained in a multifractal measure using a 
thermodynamic formalism. 

In this paper, we seek to get a characterisation of a chaotic attractor 
in terms of the underlying process that generates it. It appears that the process of 
generation of a multifractal chaotic attractor is similar to that of a typical 
Cantor set (where measure reduces after each step), with the \emph {dissipation} 
in the system playing a major role. We show this specifically below using  the 
example of Cat map which is area preserving. But a key difference is that, 
for chaotic attractors, the nature of this reduction is governed by the dynamics of the 
system. This implies that if the $D_q$ and $f(\alpha)$ curves of a chaotic attractor 
are mapped onto that of a model multiplicative process, one can derive information 
about the underlying process that generates the strange attractor, provided the 
mapping is correct. Here we try to implement this idea using 
an algorithmic scheme and show that this gives a set of parameters that can be 
used to characterise a given attractor.

A similar idea to extract the underlying multiplicative process from a multifractal 
has been applied earlier by Chhabra et.al \cite {chh1}. In order to make this inversion 
process successful, one needs to take into account two aspects, namely, the type of 
process \cite {chh1} (whether L process, P process or LP process) and the number of 
scales involved (whether two scale or multi scale). Chhabra et.al \cite {chh1} have 
shown that different multiplicative processes with only three independant parameters 
produce good fits to many of the observed $D_q$ curves. Thus the extraction of 
underlying multiplicative process, based solely on the information of $D_q$ curve, 
is nonunique and additional thermodynamic information is needed for the inversion 
process.

But the problem that we address here is slightly different. In our case, the model 
multiplicative process is fixed as a general two scale Cantor set which is the simplest 
nontrivial process giving rise to a multifractal measure. We then scan the whole set 
of parameters possible for this process (which include the L process, P process and 
LP process) and choose the statistically best fit $D_q$ 
curve to the $D_q$ spectrum computed for the attractor from the time series, which is 
then used to compute the final  $f(\alpha)$ spectrum. In this way, 
the $f(\alpha)$ spectrum of a chaotic attractor gets mapped onto that of a general 
two scale Cantor set. We show that the mapping can be done  for a large number of 
standard chaotic attractors. 
The resulting parameters can be considered to be unique to the underlying process that 
generates the attractor, upto an ambiguity regarding the number of scales involved.

The success of this procedure also implies that the $D_q$ and $f(\alpha)$ spectrum of a 
multiplicative process involving more than two scales also can be mapped onto that of a 
two scale Cantor set. We prove this theoretically as well as 
numerically in Sec.IV, by taking Cantor sets with more than 
two scales. This, in turn, suggests that though the $f(\alpha)$ spectrum has 
contributions from all the scales involved in the generation of a multifractal, 
the information contained in an $f(\alpha)$ spectrum is limited only upto two 
scales. In other words, given an $f(\alpha)$ spectrum, one can retrieve only 
the equivalent two scales which are different from the actual scales.  
Thus, while Chhabra et.al 
\cite {chh1} argues that additional information is needed to extract the underlying 
multiplicative process, our result indicate that the $f(\alpha)$ formalism itself is 
unable to extract more than two scales.

The motivation for using a Cantor set  to characterise the multifractal 
structure of a chaotic attractor comes from the fact that some well known chaotic 
attractors are believed to have underlying Cantor set structure. For example, it has been 
shown \cite {spa} that in the $x-y$ plane corresponding to $z = (r-1)$ of the 
Lorenz attractor, a transverse cut gives a multi fractal with Cantor set structure. 
Even the chaotic attractor resulting from the experimental Rayleigh-Bernard convection 
holds a support whose transverse structure is a Cantor set \cite {jen}. These Cantor 
sets are known to be characteristic of the underlying dynamics that generate the 
attractor.
 
\begin{figure*}
%\centering
\includegraphics[width=0.9\columnwidth]{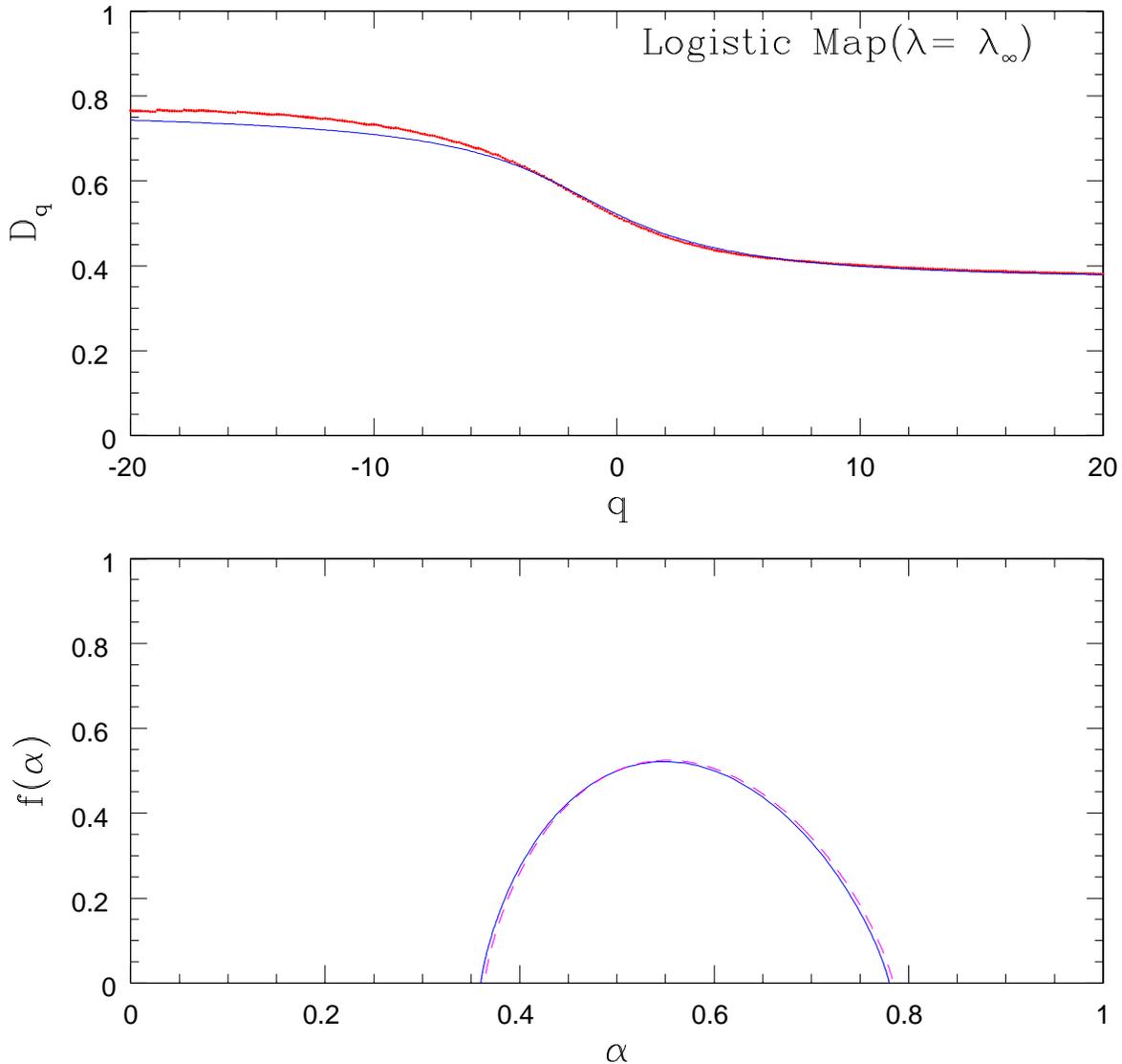}%
\caption{\label{f.1}The upper panel shows the $D_q$ values (points) and its best fit curve of the 
strange attractor at the period doubling accumulation point of the logistic map, computed from 
a time series with 30000 data points. The lower panel shows the $f(\alpha)$ spectrum computed from 
the best fit curve (dashed line) along with the theoretically predicted curve (continuous line). 
The agreement between the two is evident. }
\label{f.1}
\end{figure*}

A more general arguement to support the above statement is by using the concept of 
Kolmogorov entropy. Kolmogorov entropy can be obtained by a successively fine partition 
of the attractor in a hierarchical fashion. Going from one partition to the next 
gives one set of scales as shown in \cite {gup}. These can be treated as scales of 
higher dimensional Cantor sets. In general, there can be several scales. But the 
$f(\alpha)$ curve appears to be determined by only two scales.

In order to implement our idea, the first step is to compute the $D_q$ spectrum 
of the chaotic attractor from its time series. This is done by the standard 
delay embedding technique \cite {gra2}, but by extending the nonsubjective scheme 
recently proposed by us \cite {kph} for computing $D_2$. The $D_q$ spectrum is then 
fitted by a smooth $D_q$ curve obtained by inverse Legendre transformation equations 
\cite {atm,gra1} of the $f(\alpha)$ curve for a general two scale Cantor set. The 
statistically best fit curve is chosen by changing the parameters of the fit from 
which, the $f(\alpha)$ curve for the time series is evaluated along with a set of 
independant parameters characteristic of the Cantor set. This procedure also gives a 
couple of other interesting results. For example, we are able to propose a 
generalisation of the standard equations of two scale Cantor set for higher 
dimensions. Moreover, we explicitely derive the equations for $D_q$ and 
$f(\alpha)$ spectrum of a three scale Cantor set and show that they can be exactly 
mapped onto that of a two scale Cantor set.

Our paper is organised as follows: The details of our computational scheme are 
presented in  Sec.II and 
it is tested using time series from the logistic map and different Cantor sets 
with known parameters in  Sec.III. In Sec.IV, the $f(\alpha)$ spectrum of Cantor 
sets with more than two scales is considered both theoretically and numerically. 
Sec.V is concerned with the application of the scheme to standard chaotic attractors in 
higher dimensions. The conclusions are drawn in  Sec.VI.

\section{\label{sec:level1}NUMERICAL SCHEME}

As the first step, the spectrum of generalised dimensions $D_q$ are 
computed from the time series using the delay embedding technique 
\cite{gra2}. For a given embedding dimension $M$, the $D_q$ spectrum are 
given by the standard equation

\begin{equation}
    \label{e.1}
    D_{q} \equiv   \frac {1}{q-1} \; \lim_{R \rightarrow 0} \frac{\log \; C_q (R)}{\log \; R}
\end{equation}
where  $C_{q} (R)$ represents the generalised correlation sum.
In practical considerations, $D_q$ is computed by taking the slope of  $ \log C_{q} (R)$ versus  
$\log R$  over a scaling region. In our scheme, the scaling region is computed algorithmically 
\cite {kph} for each $D_q$ using conditions for  $R_{min}$ and  $R_{max}$ and the spectrum of 
$D_q$ for $q$ in the range $[-20,20]$ is evaluated with an error bar.

Assuming that the corresponding $f(\alpha)$ curve is a smooth convex function, we seek to 
represent it using the standard equations \cite {hal3,amr} of $\alpha$ and $f(\alpha)$ for the 
general two scale Cantor set
\begin{equation}
  \label{e.2}
  \alpha = {{r \log p_1 + (1-r) \log p_2} \over {r \log l_1 + (1-r) \log l_2}}
\end{equation}
\begin{equation}
  \label{e.3}
  f = {{r \log r + (1-r) \log (1-r)} \over {r \log l_1 + (1-r) \log l_2}}
\end{equation}
where $l_1$ and $l_2$ are the rescaling parameters and $p_1$ and $p_2$ are the 
probability measures with $p_2 = (1-p_1)$. Thus there are three independent parameters 
which are characteristic of the multiplicative process generating a given $f(\alpha)$ 
curve. Here $r$ is a variable in the range $[0,1]$, with $r \rightarrow 0$ corresponding to 
one extreme of scaling and $r \rightarrow 1$ corresponding to the other extreme. 
Taking ${{\log p_2} /{\log l_2}} > {{\log p_1} /{\log l_1}}$, as $r \rightarrow 0$, we get
\begin{equation}
  \label{e.4}
  \alpha \rightarrow \alpha_{max} \equiv {{\log p_2} \over {\log l_2}}
\end{equation}
and as $r \rightarrow 1$ 
\begin{equation}
  \label{e.5}
  \alpha \rightarrow \alpha_{min} \equiv {{\log p_1} \over {\log l_1}}
\end{equation}
By inverting Eqs.~(\ref{e.2}) and ~(\ref{e.3}) and using the standard Legendre 
transformation equations \cite {atm,gra1} connecting $\alpha$ and $f(\alpha)$ with 
$q$ and $D_q$, we get
\begin{equation}
   \label{e.6}
   q = {d \over {d\alpha}}f(\alpha)
\end{equation}
\begin{equation}
  \label{e.7}
  D_q = {{{\alpha q} - {f(\alpha)}} \over {(q-1)}}
\end{equation}
Changing the variable $\eta = 1/r$, ~(\ref{e.2}) and ~(\ref{e.3}) reduce to
\begin{equation}
  \label{e.8}
  \alpha = {{\log p_1 + (\eta -1) \log p_2} \over {\log l_1 + (\eta -1) \log l_2}}
\end{equation}
\begin{equation}
  \label{e.9}
  f = {{(\eta - 1) \log (\eta - 1) - {\eta \log \eta}} \over {\log l_1 + (\eta -1) \log l_2}}
\end{equation}
Differentiating  ~(\ref{e.8}) and ~(\ref{e.9}) with respect to $\eta$ and 
combining
\begin{equation}
  \label{e.10}
  {{df} \over {d\alpha}} = {{(\log l_1 (\log (\eta -1) - \log \eta) + \log l_2 \log \eta)} \over {(\log l_1 \log p_2 - \log l_2 \log p_1)}}
\end{equation}
Using eq.~(\ref{e.6}) and changing back to variable $r$
\begin{equation}
  \label{e.11}
  q = {{df} \over {d\alpha}} = {{\log l_1 \log (1-r) - \log l_2 \log r} \over {\log l_1 \log (1-p_1) - \log l_2 \log p_1}}
\end{equation}
Eqs.~(\ref{e.11}) and ~(\ref{e.7}) give both $q$ and $D_q$ as functions of the three 
independent parameters $l_1, l_2$ and $p_1$.

For a given set of parameters, the $D_q$ curve is determined by varying $r$ in the range 
$[0,1]$ and fitted with the computed $D_q$ values from the time series. The procedure is 
repeated by changing the values of $p_1$ in the range $[0,1]$ and for each $p_1$, 
scanning the values of $l_1$ and $l_2$ with the condition that both $l_1,l_2 < 1$. A 
statistical $\chi^2$ fitting is undertaken and the best fit curve given by the 
$\chi^2$ minimum is chosen. The complete $f(\alpha)$ curve is derived from it along 
with the complete set of parameters $p_1, l_1, l_2, \alpha_{min}$ and $\alpha_{max}$, 
for a particular time series. 

\begin{figure*}
%\centering
\includegraphics[width=0.9\columnwidth]{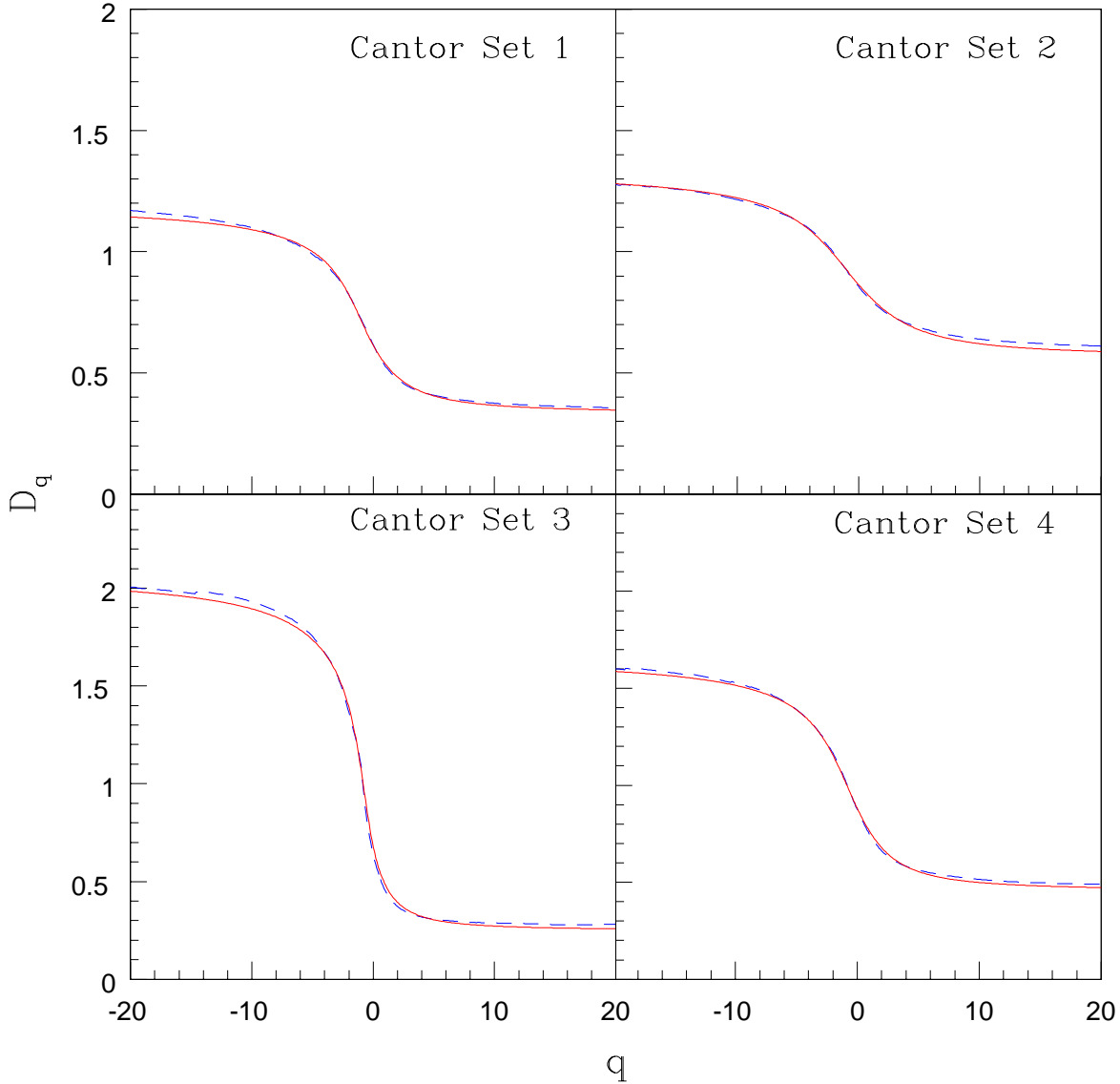}%
\caption{\label{f.2}The $D_q$ values (dashed lines) of four different two scale Cantor 
sets computed 
from their time series along with the numerical fit (continuous line) in each case.}
\label{f.2}
\end{figure*}

\begin{figure*}
%\centering
\includegraphics[width=0.9\columnwidth]{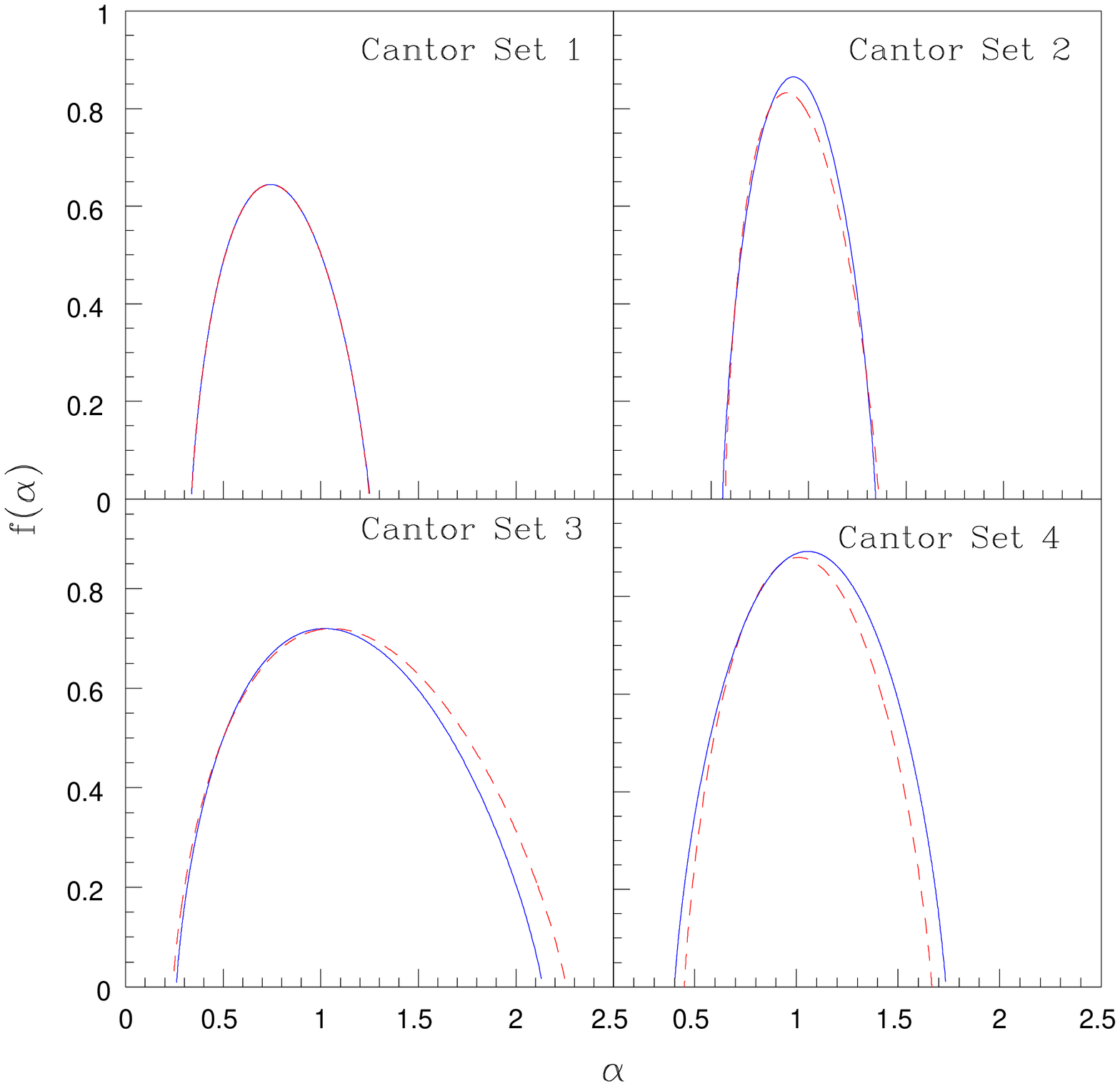}%
\caption{\label{f.3}The $f(\alpha)$ curves for the Cantor sets shown in Fig.~\ref{f.2} 
computed from the best fit $D_q$ curve (dashed line) along with the 
theoretical curves (continuous line). The parameter values  used for constructing 
the Cantor sets agree very well with those derived using our scheme in all the 
cases (see Table ~\ref{t.1}).}
\label{f.3}
\end{figure*}

\section{\label{sec:level1}TESTING THE SCHEME}

In order to illustrate our scheme, we first apply it on standard multifractals where 
the $f(\alpha)$ curve and the associated parameters are known exactly. In all the 
examples discussed in this paper, $30000$ data points are used for the analysis. The 
first one is the time series from the logistic map at the period doubling 
accumulation point. The $D_q$ spectrum is first computed using Eq.~(\ref{e.1}) (with 
$M=1$), for $q$ values in the range $[-20,+20]$. The computation is done taking a 
step width of $\Delta q = 0.1$. Choosing $p_1 = 0.5, \alpha_{min} = D_{20}$ and 
$\alpha_{max} = D_{-20}$ as input parameters, the $D_q$ curve is computed from the 
above set of equations and fitted with the $D_q$ values. The procedure is repeated by 
scanning $p_1$ in the range $[0,1]$ in steps of $0.01$. For each $p_1$, 
$\alpha_{min}$ and $\alpha_{max}$ (which in turn determine $l_1$ and $l_2$) are also 
varied independantly over a small range. The best fit $D_q$ curve is chosen as indicated 
by the $\chi^2$ minimum. Since the error in $D_q$ generally bulges as $q \rightarrow -20$, 
the error bar is also taken care of in the fitting process. The $D_q$ values computed 
from the time series and its best fit curve are shown in Fig.~\ref{f.1}.

The complete $f(\alpha)$ spectrum for the time series is computed from the best fit 
$D_q$ curve. To make a comparison, the spectrum is also determined from 
Eqs.~(\ref{e.2}) and ~(\ref{e.3}) using the known values of $p_1, l_1$ and $l_2$ for 
the logistic map, namely, $p_1 = 0.5$, $l_1 = 0.158(1/{\alpha_{F}^2})$ and 
$l_2 = 0.404(1/{\alpha_{F}})$ where $\alpha_{F}$ is Feigenbaum's universal number. Both 
the curves are also shown in  Fig.~\ref{f.1}. The three parameters derived 
using our scheme are $p_1 = 0.5, l_1 = 0.146$ and $l_2 = 0.416$ which are reasonably 
accurate considering the finiteness of the data set.

\begin{table*}
\caption{\label{t.1}Comparison of the parameters used 
for the generation of the different Cantor sets  discussed in the text 
with those computed by applying our numerical scheme. Close to 30000 points are used for 
computation in all cases.}
%\begin{center}
\begin{ruledtabular}
\begin{tabular}{lcr}
\hline
\emph{Cantor set No.}  &  Parameters used   &    Parameters computed  \\
\hline

Cantor set 1      & $p_1 = 0.60$, $l_1 = 0.22$, $l_2 = 0.48$  & $p_1 = 0.58$, $l_1 = 0.21$, $l_2 = 0.49$ \\ 

& & \\

Cantor set 2      & $p_1 = 0.42$, $l_1 = 0.22$, $l_2 = 0.67$  & $p_1 = 0.45$, $l_1 = 0.24$, $l_2 = 0.67$ \\ 

& & \\

Cantor set 3      & $p_1 = 0.66$, $l_1 = 0.18$, $l_2 = 0.62$  & $p_1 = 0.69$, $l_1 = 0.19$, $l_2 = 0.64$ \\

& & \\

Cantor set 4      & $p_1 = 0.72$, $l_1 = 0.44$, $l_2 = 0.48$  & $p_1 = 0.66$, $l_1 = 0.39$, $l_2 = 0.52$ \\

& & \\

3Scale Cantor set & $p_1 = 0.25$, $p_2 = 0.35$, $p_3 = 0.4$  &                                    \\
                  & $l_1 = 0.12$, $l_2 = 0.35$, $l_3 = 0.18$  & $p_1 = 0.50$, $l_1 = 0.26$, $l_2 = 0.52$ \\

& & \\

4Scale Cantor set & $p_1 = 0.34$, $p_2 = 0.38$, $p_3 = 0.16$, $p_4 = 0.12$  &                                    \\
                  & $l_1 = 0.12$, $l_2 = 0.25$, $l_3 = 0.18$, $l_4 = 0.08$  & $p_1 = 0.58$, $l_1 = 0.30$, $l_2 = 0.57$ \\
\hline
\end{tabular}
\end{ruledtabular}
%\end{center}
\end{table*}

As the second example, we generate time series from four  Cantor sets using  
four different sets of parameters as given in Table ~\ref{t.1}. 
Fig.~\ref{f.2} shows the computed $D_q$ values along with the best fit 
curves in all the four cases. Note that the fit is extremely accurate for the whole 
range of $q$ in all cases. The corresponding $f(\alpha)$ curves, both theoretical and 
computed from scheme are shown in Fig.~\ref{f.3}. 
The  parameter values derived from our scheme in the four cases are  
also given in Table ~\ref{t.1} for comparison. 
It is clear that the scheme recovers the complete $f(\alpha)$ spectrum and the 
parameters reasonably well. 
In order to convince ourselves that the scheme does not produce 
any spurious effects, we have also applied it to a time series from a pure white 
noise. The $D_q$ versus $q$ curve for white noise should be a straight line parallel 
to the $q$ axis with $D_0 = M$. The corresponding $f(\alpha)$ spectrum 
would be a $\delta$ function which has been verified numerically.

\begin{figure*}
%\centering
\includegraphics[width=0.9\columnwidth]{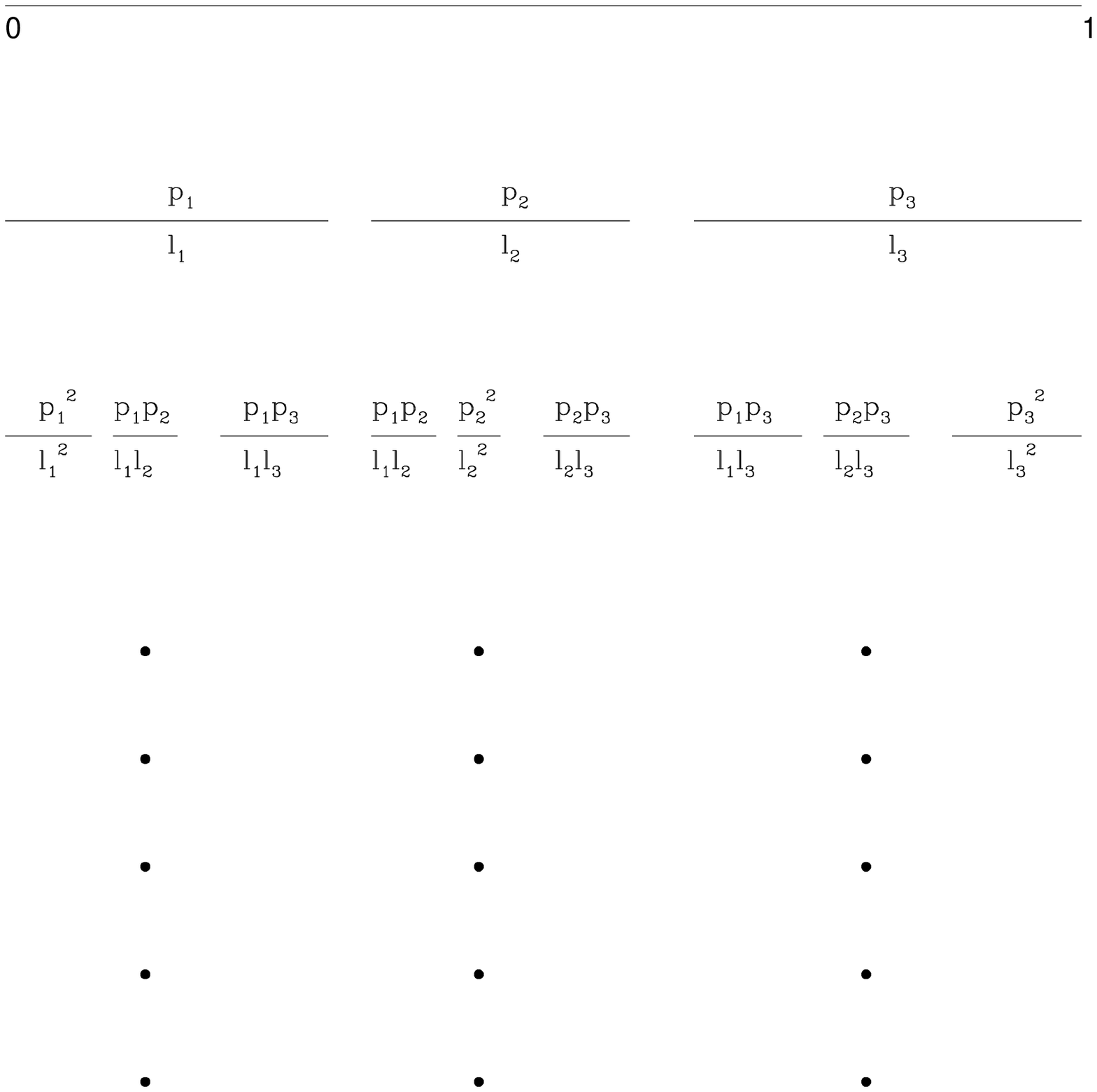}%
\caption{\label{f.4}The construction of a general 3 scale Cantor set.}
\label{f.4}
\end{figure*}

From the numerical computations of two scale Cantor sets, we also find the following 
results: While the end points of the spectrum, $\alpha_{min}$ and $\alpha_{max}$, 
are determined by the ratios $\log {p_1}/\log {l_1}$ and $\log {p_2}/\log {l_2}$,   
the peak value $D_0$ is determined by only the rescaling parameters $l_1$ 
and $l_2$. As $(l_1 + l_2)$ increases (that is, as the gap length decreases), 
$D_0$ also increases and $D_0 \rightarrow 1$ as $(l_1 + l_2) \rightarrow 1$. In 
this sense, the gap length also influences the $f(\alpha)$ spectrum indirectly. 
We will show below that this is not true in the case of three scale Cantor set 
where we miss some information regarding the scales. We also find that as the 
difference between $\alpha_{min}$ and $\alpha_{max}$ increases (that is, as the 
spectrum widens), more number of data points are required, in general, to get 
good agreement between theoretical and numerical $f(\alpha)$ curves.

\begin{figure*}
%\centering
\includegraphics[width=0.9\columnwidth]{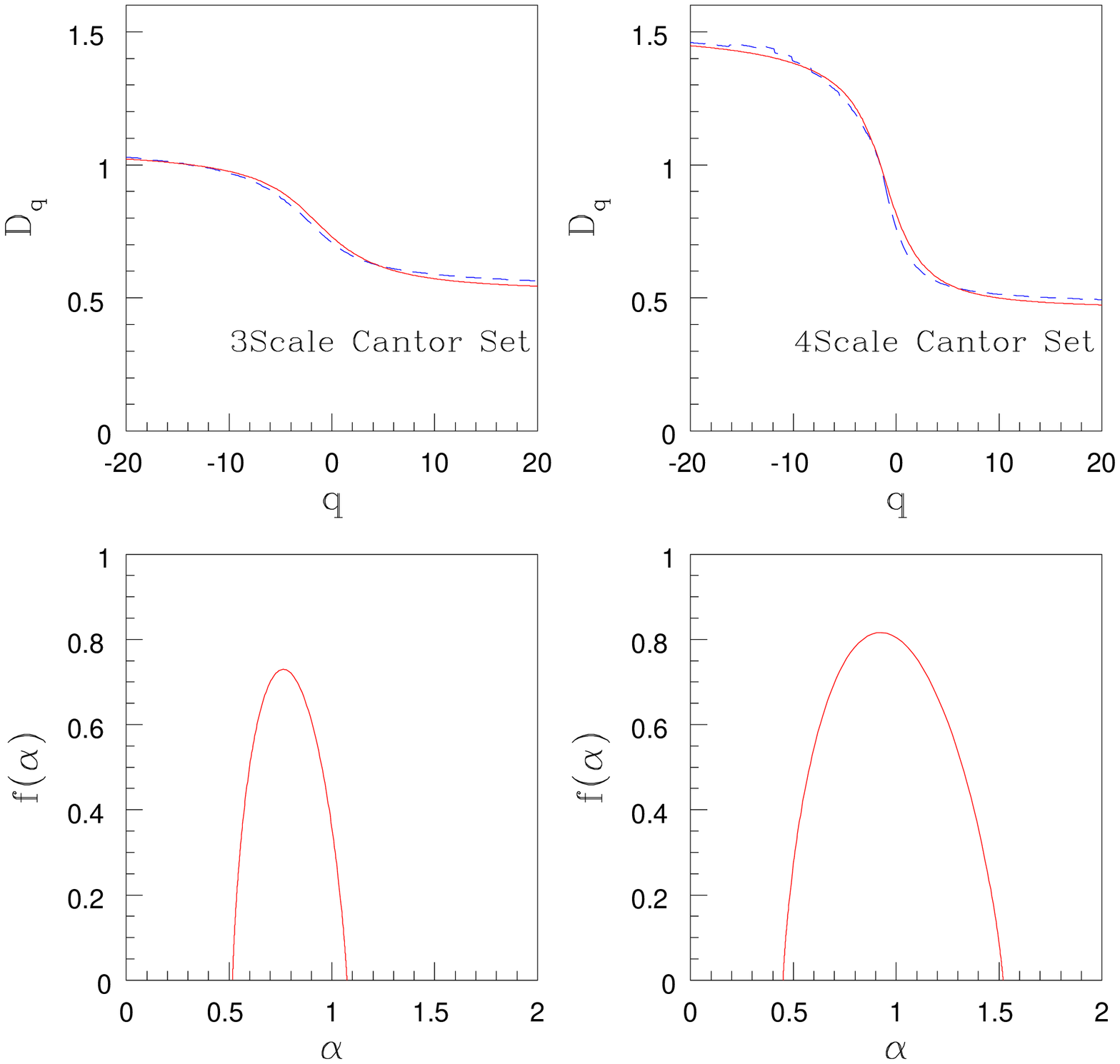}%
\caption{\label{f.5}The $D_q$ and $f(\alpha)$ curves of a 3 scale and a 4 scale 
Cantor set computed using our scheme. The $D_q$ curves computed from the time series 
(dashed lines) are fitted with that of equivalent two scale Cantor sets 
(continuous lines) in 
both cases. The parameters used and computed by the scheme are compared in 
Table ~\ref{t.1}.}
\label{f.5}
\end{figure*}

\begin{figure*}
%\centering
\includegraphics[width=0.9\columnwidth]{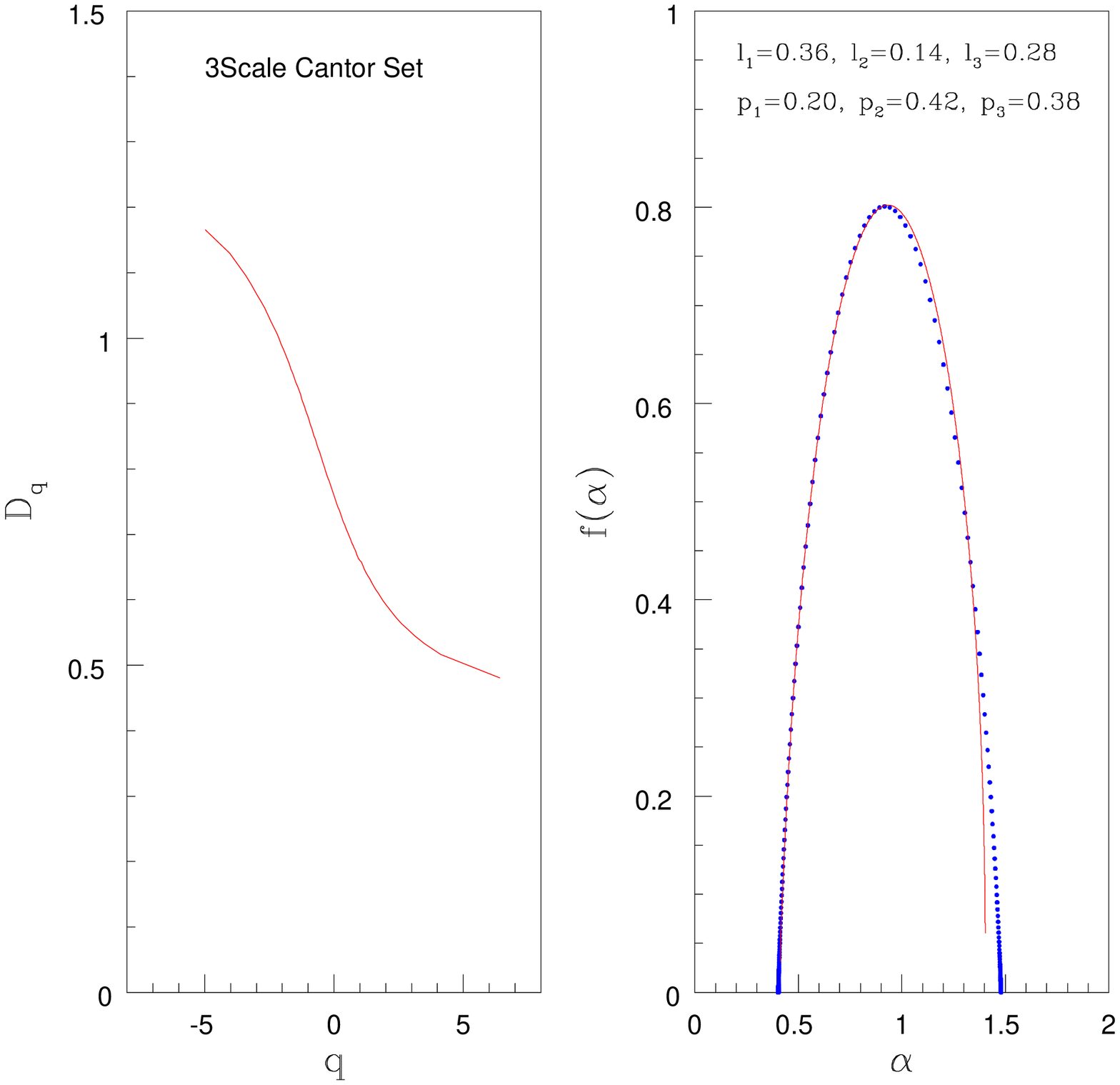}%
\caption{\label{f.6}The left panel shows the theoretical $D_q$ curve of a 3 scale 
Cantor set with parameters shown in the figure. The right panel shows the corresponding 
$f(\alpha)$ curve (continuous line) calculated theoretically using 
Eqs. ~(\ref{e.22}) and ~(\ref{e.24}). This is superimposed on the $f(\alpha)$ 
spectrum (points) computed using our numerical code by fitting 2 scale Cantor set. 
Close to 30000 data points are used for the computations.}
\label{f.6}
\end{figure*}

\section{\label{sec:level1}MULTI SCALE CANTOR SETS}

In this section, we consider the $f(\alpha)$ spectrum of a Cantor set with more than two 
scales. First we show the numerical results using our scheme.  For this, we first 
generate the time series for a general 3 scale Cantor set  and compute its $D_q$ spectrum. 
The geometrical construction of a 
general 3 scale Cantor set is shown in Fig.~\ref{f.4}. At every stage, an interval gets 
subdivided into three so that the set involves 3 rescaling parameters $l_1$, $l_2$, $l_3$ 
and 3 probability measures $p_1$, $p_2$, $p_3$ as shown.
The numerically computed $D_q$ spectrum  for a typical 3 scale Cantor set (with parameters 
given in  Table ~\ref{t.1}) is shown in Fig.~\ref{f.5} 
(upper left panel). The $D_q$ curve can be very well fitted by a 2 scale Cantor set and the 
complete $f(\alpha)$ spectrum for the 3 scale Cantor set is evaluated (lower left panel). 
We have repeated our computations for a 
4 scale Cantor set as well and the results are also shown in  Fig.~\ref{f.5} (right panel). 
In both cases, the parameters used for the construction of the Cantor sets and those computed 
by our scheme are given in  Table ~\ref{t.1}. Thus it is clear that the $f(\alpha)$ spectrum 
cannot pick up the full information about the various scales and probability measures. 
No matter how many scales are involved in the generation of the multifractal, the 
$f(\alpha)$ spectrum can be  reproduced by an equivalent 2 scale Cantor set.

We now derive explicite expressions for $\alpha$ and $f(\alpha)$ for a 3 scale 
Cantor set.  We follow 
the arguements given in Halsey et.al \cite {hal3}, Sec.II-C-4. For the 3 scale 
Cantor set, one can write
\begin{equation}
  \label{e.12}
  \Gamma (q,\tau,n) = \left({{p_{1}^q}\over {l_{1}^{\tau}}} + {{p_{2}^q}\over {l_{2}^{\tau}}} + {{p_{3}^q}\over {l_{3}^{\tau}}}\right)^n = 1
\end{equation}
Expanding
\begin{equation}
  \label{e.13}
  \Gamma (q,\tau,n) = \sum_{m_1,m_2} {{n!}\over {m_1!m_2!(m-m_1-m_2)!}}p_1^{m_1q}p_2^{m_2q}p_3^{(m-m_1-m_2)q}l_1^{-m_1\tau}l_2^{-m_2\tau}l_3^{-(m-m_1-m_2)\tau} = 1
\end{equation}
In the limit $n \rightarrow \infty$, the largest term contributes. Hence we have
\begin{equation}
  \label{e.14}
  {{\partial\Gamma}\over {\partial m_1}} = 0
\end{equation}
\begin{equation}
  \label{e.15}
  {{\partial\Gamma}\over {\partial m_2}} = 0
\end{equation}
Using the Stirling approximation and simplifying the above two conditions we get
\begin{equation}
  \label{e.16}
  -\log r + \log (1-r-s) + q\log (p_1/p_3) - \tau \log (l_1/l_3) = 0
\end{equation}
\begin{equation}
  \label{e.17}
   -\log s + \log (1-r-s) + q\log (p_2/p_3) - \tau \log (l_2/l_3) = 0
\end{equation}
where $r = m_1/n$ and $s = m_2/n$ are free parameters. Also from Eq.~(\ref{e.13}), using a 
similar procedure, one can  show that
\begin{equation}
  \label{e.18}
  r\log r - s\log s - (1-r-s)\log (1-r-s) + q(r \log p_1 + s\log p_2 + (1-r-s)\log p_3) - \tau (r\log l_1 + s\log l_2 + (1-r-s)\log l_3) = 0 
\end{equation}
Combining Eqs.~(\ref{e.16}), ~(\ref{e.17}) and ~(\ref{e.18}) and eliminating $\tau$  
we get the following relations for $q$
\begin{equation}
  \label{e.19}
  q = {{\log (l_2/l_3)\log ((1-r-s)/r) - \log (l_1/l_3)\log ((1-r-s)/s)}\over {\log (l_1/l_3)\log (p_2/p_3) - \log (l_2/l_3)\log (p_1/p_3)}}
\end{equation}
\begin{equation}
  \label{e.20}
  q = {{\log (l_1/l_3)(-r\log r - s\log s - (1-r-s)\log (1-r-s)) - (r\log l_1 + s\log l_2 + (1-r-s)\log l_3)\log ((1-r-s)/r)}\over {(r\log l_1 + s\log l_2 + (1-r-s)\log l_3)\log (p_1/p_3) - \log (l_1/l_3)(r\log p_1 + s\log p_2 + (1-r-s)\log p_3)}}
\end{equation}
These two equations for $q$ can be used to obtain a relation between $r$ and $s$.

To compute the $D_q$ spectrum, vary $r$ from 0 to 1. For every value of $r$, the value of $s$ 
that satisfies the Eqs.~(\ref{e.19}) and ~(\ref{e.20})simultaneously  is 
computed numerically, with the condition that $0 < s < (1-r)$. For every value of $r$ and 
$s$, $q$ and $\tau$ can be determined using Eqs.~(\ref{e.19}) and ~(\ref{e.16}), 
which in turn gives $D_q = \tau /(q-1)$.

The singularity exponent $\alpha$ is determined by the condition
\begin{equation}
  \label{e.21}
  p_1^{m_1}p_2^{m_2}p_3^{(m-m_1-m_2)} = \left(l_1^{m_1}l_2^{m_2}l_3^{(m-m_1-m_2)}\right)^{\alpha}
\end{equation}
This gives the expression for $\alpha$
\begin{equation}
  \label{e.22}
  \alpha = {{r\log p_1 + s\log p_2 + (1-r-s)\log p_3}\over {r\log l_1 + s\log l_2 + (1-r-s)\log l_3}}
\end{equation}
Similarly, the density exponent $f(\alpha)$ is determined by
\begin{equation}
  \label{e.23}
  n!\left(l_1^{m_1}l_2^{m_2}l_3^{(m-m_1-m_2)}\right)^{f(\alpha)} = m_1!m_2!(m-m_1-m_2)!
\end{equation}
which gives the following expression for $f(\alpha)$
\begin{equation}
  \label{e.24}
  f(\alpha) = {{r\log r + s\log s + (1-r-s)\log (1-r-s)}\over {r\log l_1 + s\log l_2 + (1-r-s)\log l_3}}
\end{equation}
By varying $r$ from 0 to 1, the $f(\alpha)$ spectrum for a given 3 scale Cantor set 
can be determined theoretically. In Fig.~\ref{f.6}, the theoretically computed $D_q$ 
and $f(\alpha)$ spectrum for a typical 3 scale Cantor set is shown. Along with the 
theoretical $f(\alpha)$ curve, we also show the numerical one (points) for the 
same Cantor set, computed using our scheme. Thus it is evident that the $f(\alpha)$ 
spectrum of a 3 scale Cantor set can be mapped  onto that of a 2 scale 
Cantor set. Also, our numerical results on 4 scale Cantor set (Fig.~\ref{f.5}) suggests that 
this mapping onto 2 scale Cantor set can possibly be extended for 
the $f(\alpha)$ spectrum of four or more scale Cantor sets.

\begin{figure*}
%\centering
\includegraphics[width=0.9\columnwidth]{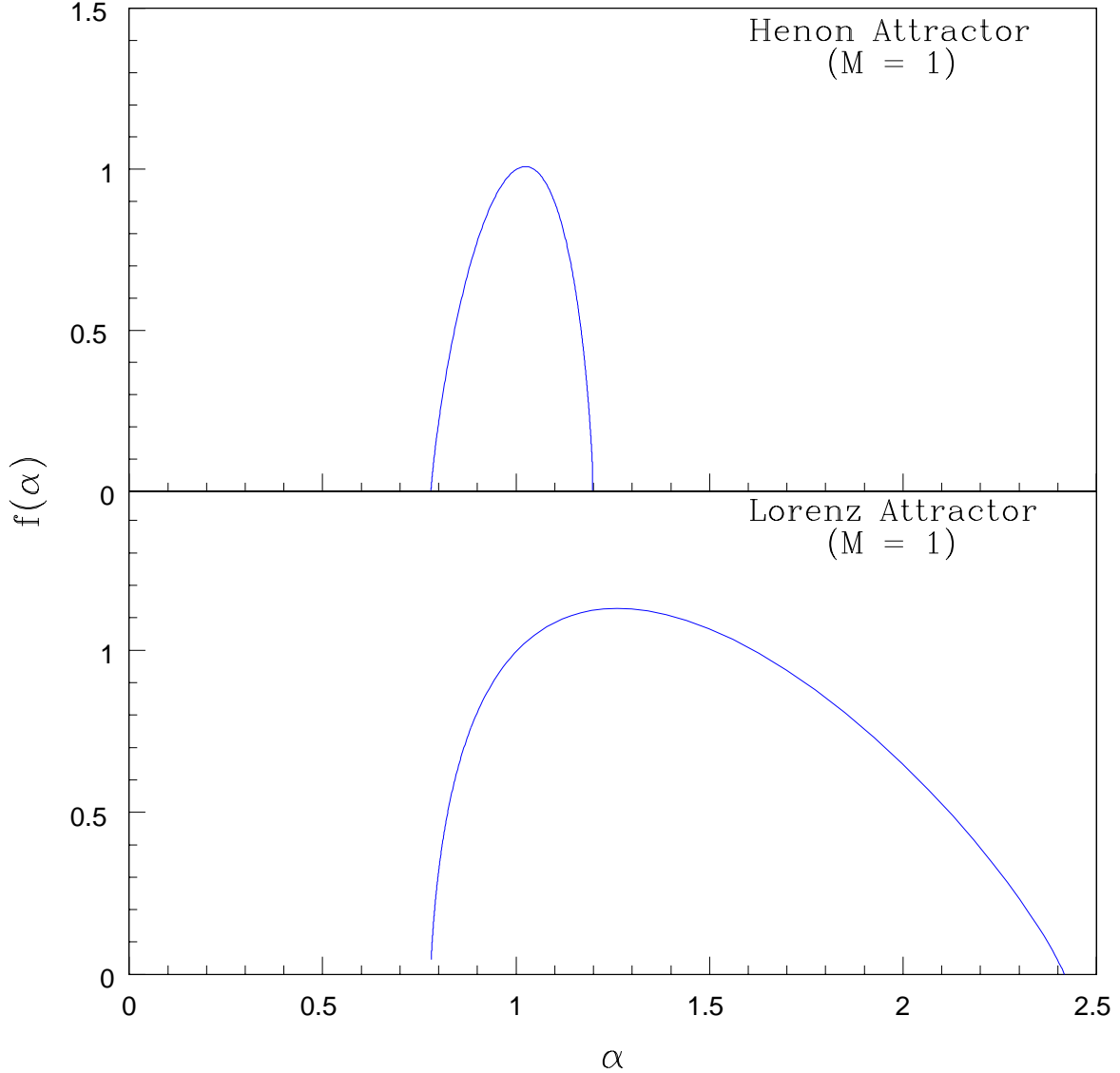}%
\caption{\label{f.7}The multifractal spectrum of the one dimensional 
projection of Henon and Lorenz attractors, which show the underlying 
Cantor set structure. Note that the peak values of the spectrum in both 
cases are equal to 1.}
\label{f.7}
\end{figure*}

\begin{figure*}
%\centering
\includegraphics[width=0.9\columnwidth]{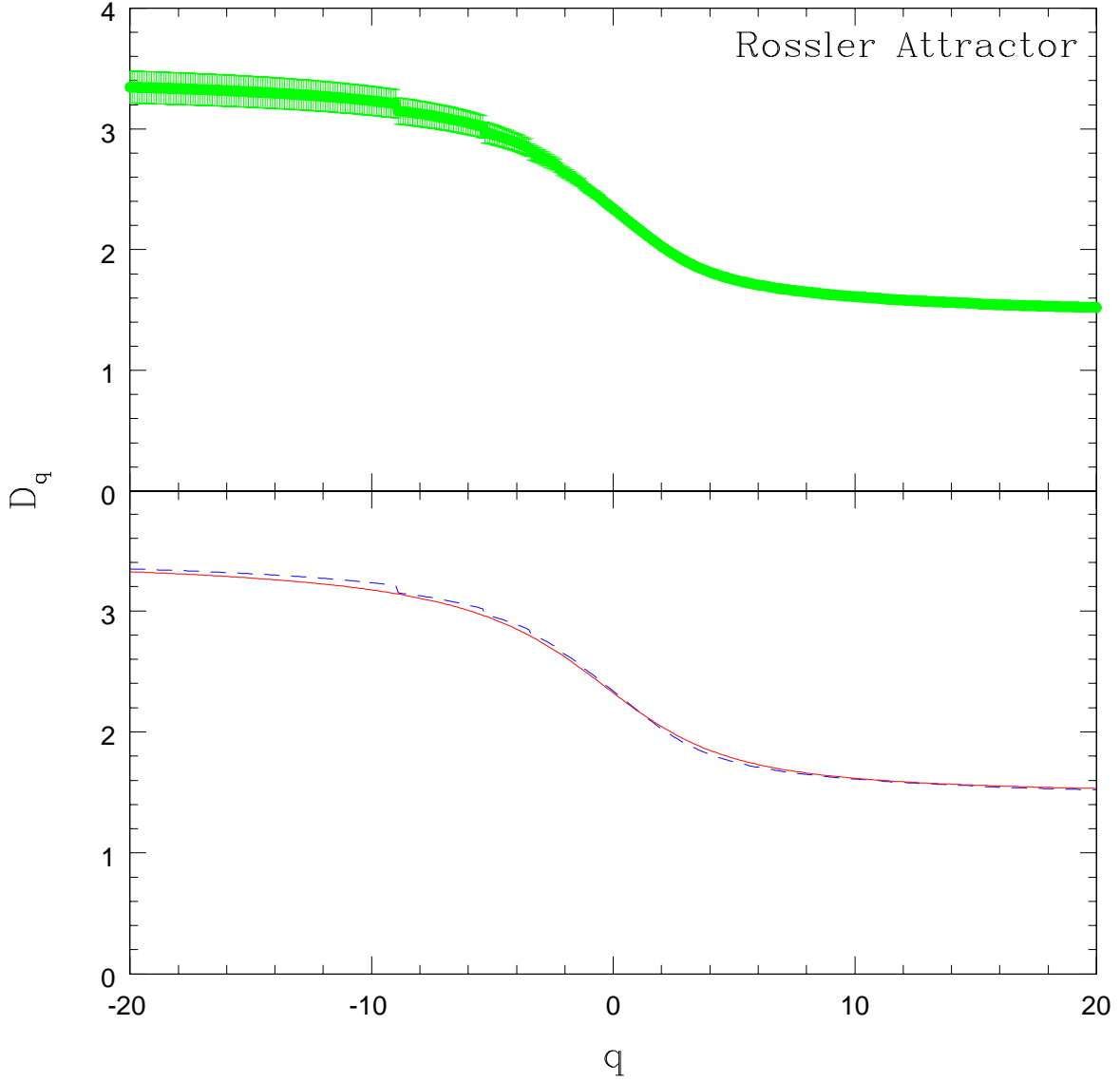}%
\caption{\label{f.8}The upper panel gives the $D_q$ spectrum, with error bar, of the 
Rossler attractor 
computed from the time series. The lower panel shows the accuracy of fitting with 
again the $D_q$ values (dashed line) and its best fit curve (continuous line) computed 
using the scheme.}
\label{f.8}
\end{figure*}

\begin{figure*}
%\centering
\includegraphics[width=0.9\columnwidth]{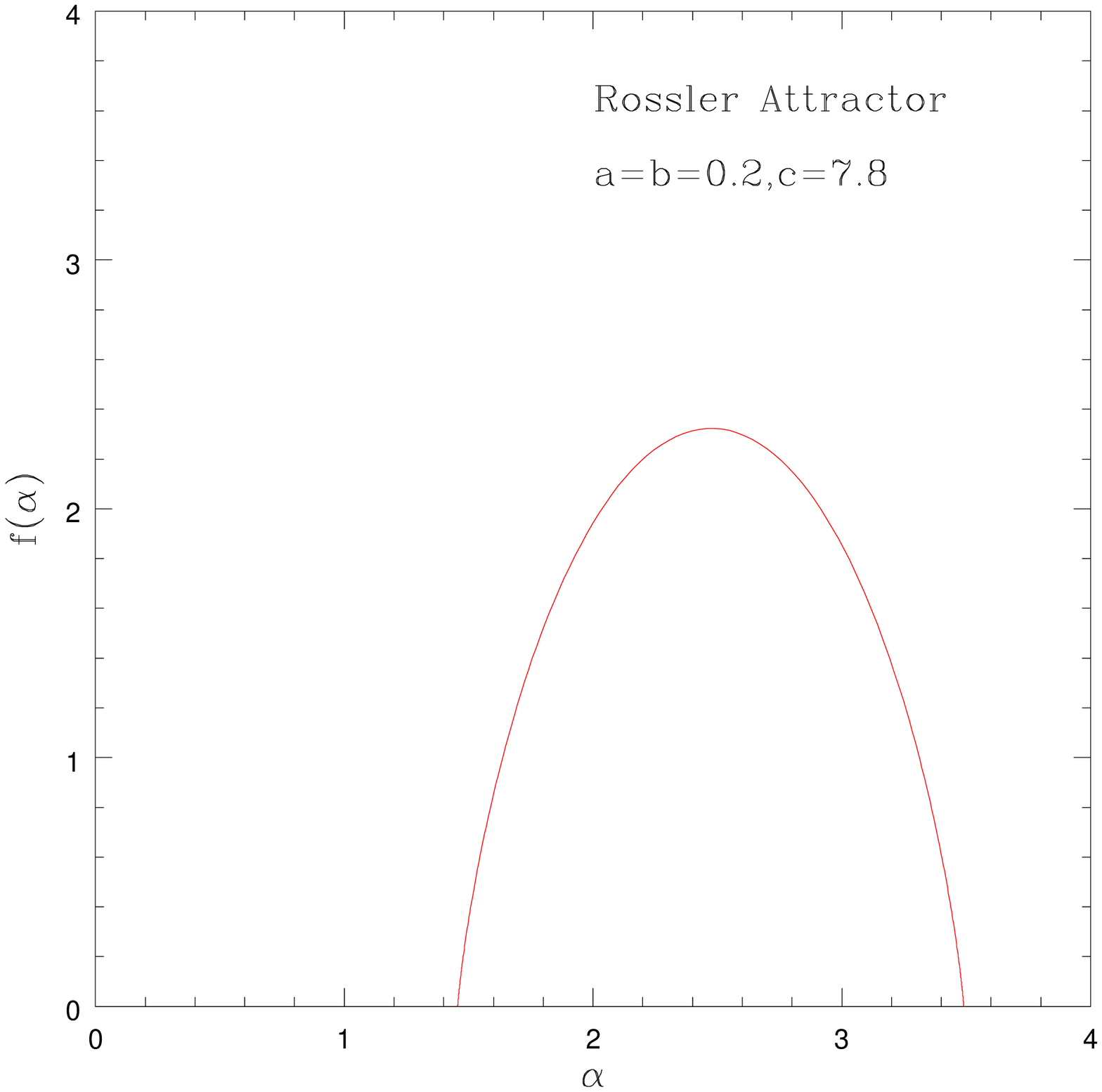}%
\caption{\label{f.9}The $f(\alpha)$ curve of the Rossler attractor computed from the 
best fit $D_q$ curve shown in Fig.~\ref{f.8}.}
\label{f.9}
\end{figure*}

\section{\label{sec:level1}CHARACTERISATION OF STRANGE ATTRACTORS}

Evaluating the $f(\alpha)$ spectrum of one dimensional sets is straightforward. But 
computing the spectra of even synthetic higher dimensional attractors is a 
challenging task. Generally, the $f(\alpha)$ spectrum for higher dimensional chaotic 
attractors is 
calculated taking only one dimension \cite {elg,chh3,wik}, which 
characterise the transverse self similar structure on the attractor equivalent to 
a Cantor set. In the resulting $f(\alpha)$ spectrum, the peak value 
(that is, $D_0$) will be equal to $1$, as the higher dimensional attractor is 
projected into one dimension. This is shown in Fig.~\ref{f.7} 
for Henon and Lorenz attractors and the results are consistent with the earlier results. 

In order to extend our scheme to higher dimensional strange 
attractors, their $f(\alpha)$ spectra are to be considered analogous to a two scale 
Cantor measure in higher dimensions. While the $f(\alpha)$ curve can be recovered 
using the correct embedding dimension $M$, the meaning of the parameters have to be 
interpreted properly. For a one dimensional Cantor set, $p_1$ is a probability 
measure while $l_1$ and $l_2$ are fractional lengths at each stage. Extending this 
analogy to two and three dimensions, $p_1$ can still be interpreted as a probability 
measure for the two higher dimensional scales, say $\tau_1$ and $\tau_2$. These can be 
considered as fractional measures corresponding to area or volume depending on the 
embedding dimension $M$. In other words, $p_1$ is a measure representing the 
underlying dynamics, while $\tau_1$ and $\tau_2$ correspond to geometric scaling. This 
gives an alternate description of the formation of a strange attractor if it is 
correlated to a higher dimensional analogue of the Cantor set. 

As discussed in Sec.II, for the one dimensional Cantor set, 
$\alpha_{min}$ and 
$\alpha_{max}$ are given by Eqs.~(\ref{e.4}) and ~(\ref{e.5}) in terms of the 
parameters. For $p_1 = p_2$ and $l_1 = l_2, \alpha_{min} = \alpha_{max} \leq 1$ and 
the set becomes a simple fractal with $\alpha \equiv f(\alpha) = D_0$, the fractal 
dimension. Extending this analogy to higher dimensions, we propose that  
Eqs.~(\ref{e.4}) and ~(\ref{e.5}) are to be modified as
\begin{equation}
  \label{e.25}
  \alpha_{max} = M {{\log p_2} \over {\log \tau_2}}
\end{equation}
and 
\begin{equation}
  \label{e.26}
  \alpha_{min} = M {{\log p_1} \over {\log \tau_1}}
\end{equation}
As in the case of one dimensional Cantor sets, 
for $p_1 = p_2$ and $\tau_1 = \tau_2$, $\alpha_{max} = \alpha_{min} \leq M$ and the 
set is again a simple fractal with fractal dimension $D_0 = \alpha \equiv f(\alpha)$. 
Rewriting the above equations,
\begin{equation}
  \label{e.27}
  \alpha_{max} = {{\log p_2} \over {\log ({\tau_{2}}^{1/M})}} = {{\log p_2} \over {\log l_2}}
\end{equation}
and 
\begin{equation}
  \label{e.28}
  \alpha_{min} = {{\log p_1} \over {\log ({\tau_{1}}^{1/M})}} = {{\log p_1} \over {\log l_1}}
\end{equation}
Replacing $l_1$ and $l_2$ by $\tau_{1}^{1/M}$ and $\tau_{2}^{1/M}$ in 
Eqs.~(\ref{e.2}) and ~(\ref{e.3}), the defining equations for the two scale 
Cantor set in $M$ dimension can be generalised as 
\begin{equation}
  \label{e.29}
  \alpha = {M[{r \log p_1 + (1-r) \log p_2}] \over {r \log {\tau_1} + (1-r) \log {\tau_2}}}
\end{equation}
\begin{equation}
  \label{e.30}
  f = {M[{r \log r + (1-r) \log (1-r)}] \over {r \log {\tau_1} + (1-r) \log {\tau_2}}}
\end{equation}
Just like $l_1 + l_2 < 1$ for one dimensional Cantor set, we expect 
$\tau_1 + \tau_2 < 1$ in $M$ dimensions. This is because, the measure keeps on reducing 
after each time step due to dissipation and $\tau_1$ and $\tau_2$ represent the 
fractional reduction in the measure for the two scales. It should be noted that 
since, in general, different scales apply in different directions, $\tau_1$ and $\tau_2$ 
should be treated as some effective scales in higher dimension.

\begin{table*}
\caption{\label{t.2}The complete set of parameters computed using our scheme for 
six standard chaotic attractors. See \cite {spr}, for example, for the details of 
the chaotic systems and the values of the parameters used.}
%begin{center}
\begin{ruledtabular}
\begin{tabular}{ccccccc}
\hline
\emph{Attractor}  &  $\alpha_{min}$  &  $\alpha_{max}$  &  $D_0$            &  $p_1$  & $\tau_1$ & $\tau_2$ \\
\hline

Rossler attractor & 		     & 		        &                   &         &        &       \\
($a=b=0.2, c=7.8$) & $1.46 \pm 0.02$ & $3.39 \pm 0.14$  & $2.31 \pm 0.02$   & $0.65$  & $0.42$ & $0.41$  \\  

& & \\

Lorenz attractor     	 &		   &		     &                 &        &        &       \\
($\sigma=10,r=28,b=8/3$) & $1.38 \pm 0.03$ & $3.71 \pm 0.12$ & $2.16 \pm 0.04$ & $0.50$  & $0.22$ & $0.57$ \\ 

& & \\

Ueda attractor   &		   &		     &	               &        &         &            \\
($k=0.05,A=7.5$) & $1.73 \pm 0.05$ & $3.78 \pm 0.13$ & $2.62 \pm 0.06$ & $0.64$ & $0.46$  & $0.44$     \\ 

& & \\

Duffing attractor   	  &		    &		      &                 &        &        &   \\
($b=0.25,A=0.4,\Omega=1$) & $1.84 \pm 0.04$ & $3.59 \pm 0.08$ & $2.78 \pm 0.04$ & $0.81$ & $0.71$ & $0.25$ \\ 

& & \\

Henon attractor & 				& 	&	  &         &         &              \\
($a=1.4,b=0.3$) & $0.96 \pm 0.02$ & $2.27 \pm 0.08$ & $1.43 \pm 0.03$  & $0.50$  & $0.24$  & $0.54$    \\ 

& & \\

Tinkerbell attractor & 				& 	&	  &         &         &              \\
($a=0.9,b=-0.6,c=2,d=0.5$) & $0.83 \pm 0.02$ & $3.43 \pm 0.12$ & $1.65 \pm 0.03$ & $0.60$  & $0.29$  & $0.58$     \\ 
\hline
\end{tabular}
\end{ruledtabular}
%end{center}
\end{table*}

\begin{figure*}
%\centering
\includegraphics[width=0.9\columnwidth]{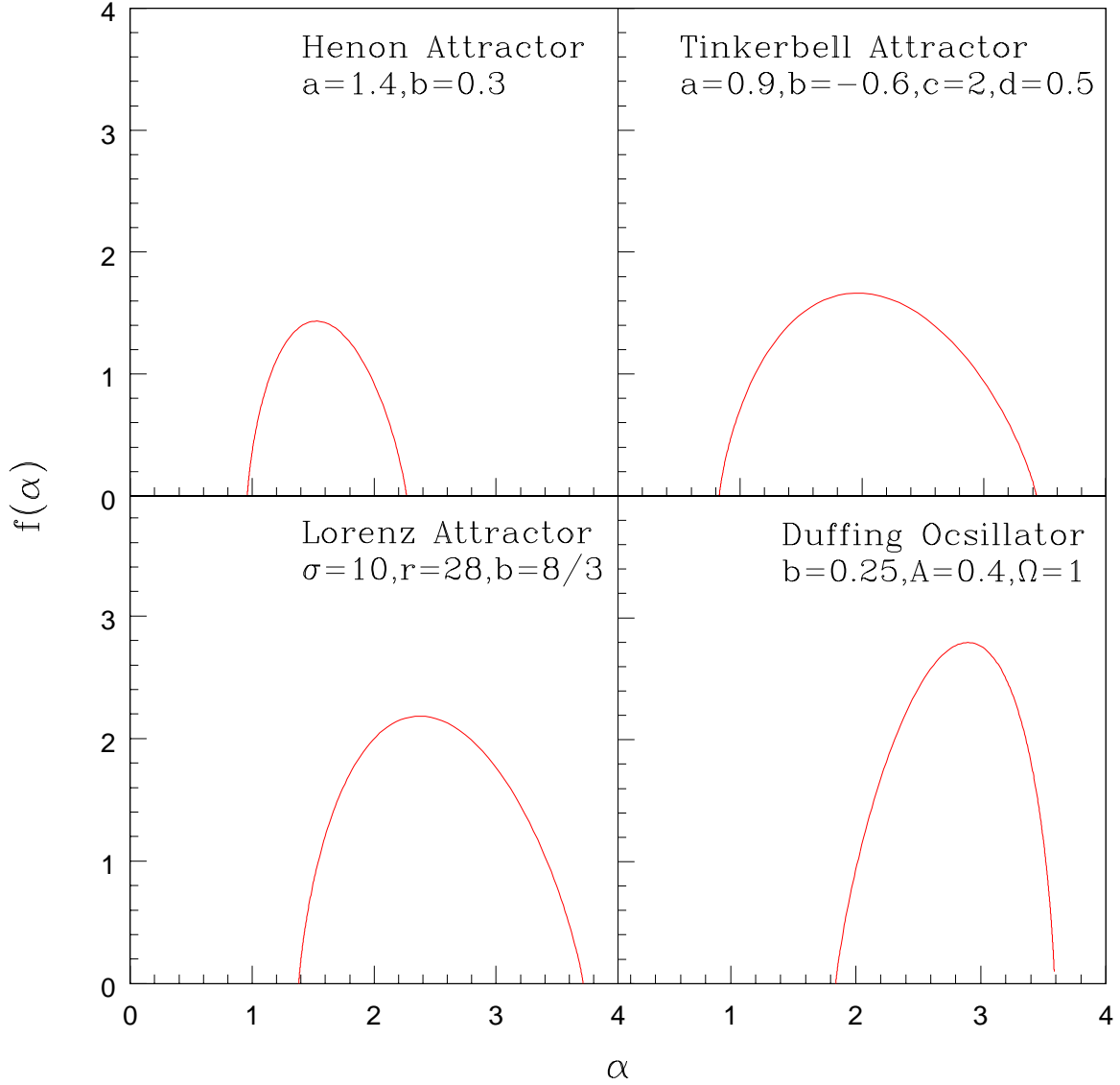}%
\caption{\label{f.10}The multifractal spectrum of four standard chaotic attractors 
computed by applying our algorithmic scheme. The parameter values used for generating 
the time series are also shown.}
\label{f.10}
\end{figure*}

\begin{figure*}
%\centering
\includegraphics[width=0.9\columnwidth]{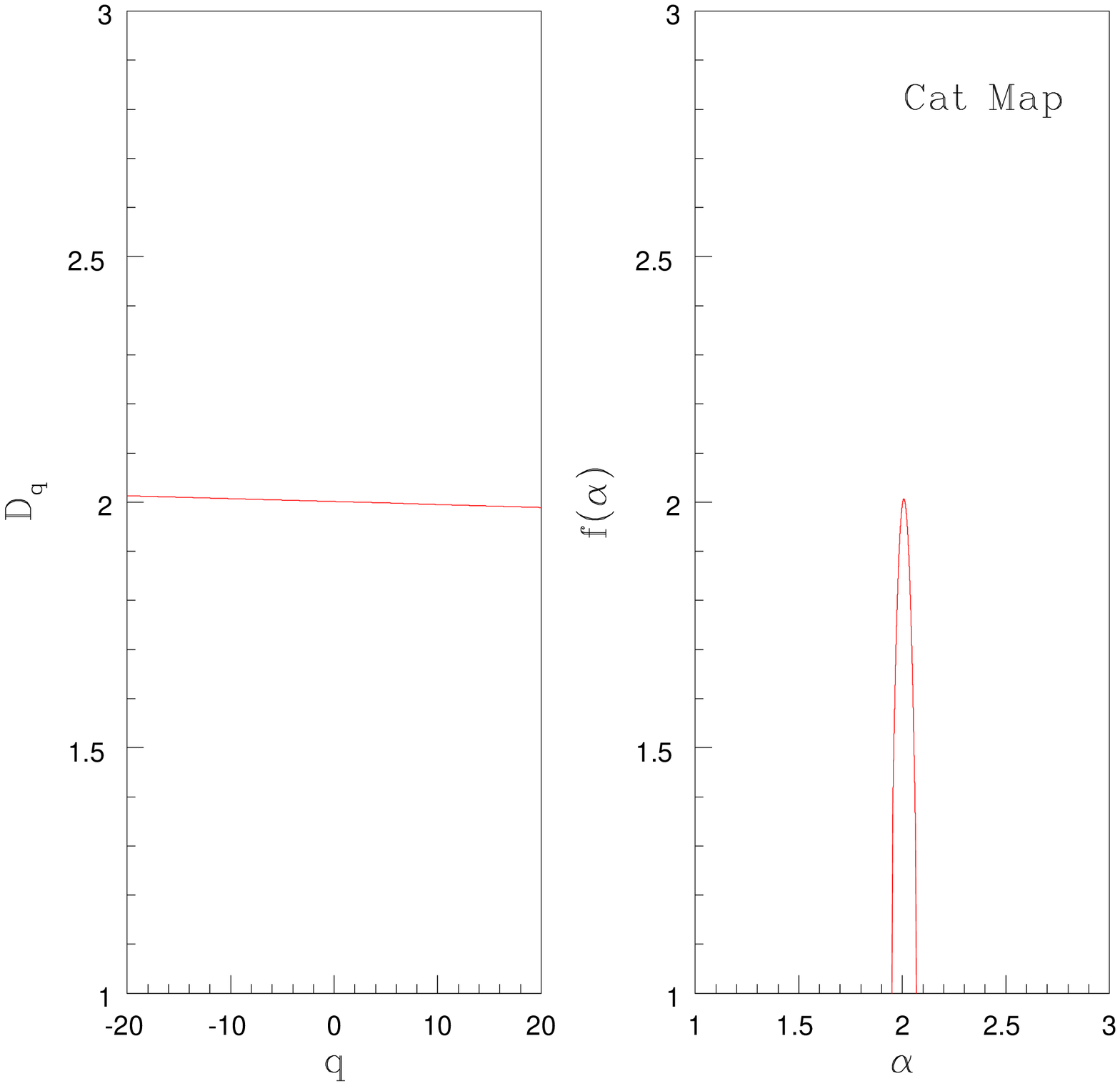}%
\caption{\label{f.11}The $D_q$ curve of the Cat map (left panel) computed using our 
scheme which is a straight line parallel to the $q$  axis just like that of a 
white noise. The corresponding $f(\alpha)$ curve is shown in the right panel, which is 
$\delta$ function with $\alpha \equiv f(\alpha) = 2$.}
\label{f.11}
\end{figure*}

We now check these results using the time series from a standard chaotic attractor, 
namely the Rossler attractor for parameter values $a = b = 0.2$ and $c = 7.8$ with 
$30000$ data points. Fig.~\ref{f.8} shows the $D_q$ spectrum  computed 
from the time series taking $M = 3$, along with the best fit curve applying our scheme. 
The fit is found to be very good for the whole range of $q$ values. The complete 
$f(\alpha)$ spectrum computed from the best fit $D_q$ curve is shown in Fig.~\ref{f.9}. 
The scheme also calculates the three parameters as $p_1 = 0.65$,  
$\tau_1 = 0.42$ and $\tau_2 = 0.41$ so that $\tau_1 + \tau_2 = 0.83 < 1$. 
Thus one can say that if the $f(\alpha)$ spectrum of the Rossler attractor is made 
equivalent to a two scale Cantor set in three dimension, the resulting probability 
measures are $0.65$ and $0.35$ and rescaling parameters $0.42$ and $0.41$. Interestingly, 
it appears that the Rossler attractor is generated by a P process rather than a 
LP process.

The scheme has also been applied to several standard chaotic attractors in two and three 
dimensions. The $f(\alpha)$ spectrum are shown in Fig.~\ref{f.10} for four of them, while 
the complete set of parameters for six standard chaotic attractors are given in 
Table ~\ref{t.2}. The error bars given for $\alpha_{max}, \alpha_{min}$ and $D_0$ are 
those reflected from the computed $D_q$ values. In a way, the two sets of parameters 
given above, that is $p_1, \tau_1, \tau_2$ and  $\alpha_{max}, \alpha_{min}, D_0$, can be 
considered as complementary to each other. While the former contain the finger 
prints of the underlying process that generate the strange attractor (the extent of 
stretching and folding and redistribution of measures at each time step), the latter 
characterises the geometric complexity of the attractor once it is formed. Both can be 
independantly used to differentiate between chaotic attractors formed from different 
systems or from the same system for different parameter values. The former may be 
more relevant in the case of chaotic attractors obtained from experimental systems.

Finally, we wish to point out that dissipation is a key factor leading to the 
multifractal nature of a chaotic attractor. To show this, we consider a counter 
example, namely, that of Cat map which is area preserving.
The fixed points of the Cat map are \emph {hyperbolic}, which are neither 
attractors nor repellers and the trajectories uniformly fill the phase space as 
time $t \rightarrow \infty$. Its $D_q$ spectrum computed from the time series 
is found to be a straight line as shown in Fig.~\ref{f.11}, just like that of a 
white noise. The corresponding  $f(\alpha)$ curve is a $\delta$ function,  
also shown in Fig.~\ref{f.11}. Since $\alpha_{min} = \alpha_{max}$, a two scale 
fit gives the parameters as $p_1 = 0.5, \tau_1 = 0.49$ and $\tau_2 = 0.51$. 
Thus the Cat map attractor turns out to be a simple fractal rather than a 
multifractal.

\section{\label{sec:level1}DISCUSSION AND CONCLUSION}
In this paper, we show that a chaotic attractor can be  characterised using a 
set of three independant parameters which are specific to the underlying process generating it. 
The method relies on a scheme that maps the $f(\alpha)$ spectrum of a chaotic attractor 
onto that of a general two scale Cantor set. The scheme is first tested using one 
dimensional chaotic attractors and Cantor sets whose $f(\alpha)$ curves and parameters 
are known  and subsequently applied to higher dimensional cases.

In the scheme, the $D_q$ spectrum of a chaotic attractor is compared with the $D_q$ 
curve computed from a model multiplicative process. Similar idea has also been used to 
deduce certain statistical characteristics of a system and infer features of the 
dynamical processes leading to the observed macroscopic parameters. One such example 
has been provided earlier by Meneveau and Sreenivasan \cite{sre2} in the study of energy 
dissipation rate in fully developed turbulent flows. By comparing the experimental 
$D_q$ data with that of a two scale Cantor measure, they have shown that the dynamics 
leading to the observed multifractal distributions of the energy dissipation rate 
can be well approximated by a single multi step process involving unequal 
energy distribution in the ratio $7/3$.

Usually, a multifractal is characterised only by the range of scaling involved 
$[\alpha_{min},\alpha_{max}]$, which roughly represents the inhomogeinity of the 
attractor. So the set of parameters computed here seems to give alternative way of 
characterising them. But we wish to emphasize that the information contained in these 
parameters is more subtle. For example, once these prameters are known, $\alpha_{min}$ and 
$\alpha_{max}$ can be determined using Eqs.~(\ref{e.27}) and ~(\ref{e.28}). Thus by 
evaluating  $p_1$, $l_1$ and $l_2$, we get additional information regarding the 
dynamic process leading to the generation of the strange attractor. Moreover, these 
parameters can also give indication as to \emph {why} the degree of inhomogeinity varies 
between different chaotic attractors. As is well known from the srudy of Cantor sets, 
the primary reason for the increased inhomogeinity is the wide difference between the 
rescaling measures $l_1$ and $l_2$. Looking at the parameter values, rescaling 
measures $\tau_1$ and $\tau_2$ are very close for Rossler and Ueda attractors which 
appear less inhomogeneous, while that for Lorenz and Duffing are widely different 
making them more inhomogeneous with two clear scrolls.

Another novel aspect of the scheme worth commenting is the use of two scale Cantor 
measures in higher dimension as analogues of chaotic attractors. Eventhough such 
objects are not much discussed in the literature, one can envisage them, for example, 
as generalisation of the well known Sierpinsky carpets in two dimension or the 
Menger sponge in three dimension. But a key difference between these and a chaotic 
attractor is that the rescaled measures are not regular in the generation of the latter. 
Recently, Perfect et. al \cite {per} present a 
general theoretical framework for generating geometrical multifractal Sierpinsky carpets 
using a generator with variable mass fractions determined by the truncated binomial 
probability distribution and to compute their generalised dimensions. 
It turns out that  the chaotic attractors are more similar to multifractals generated 
in higher 
dimensional support, such as, fractal growth patterns and since the rescaled 
measures are irregular, a one dimensional measure such as $l_1 = \tau_{1}^{1/M}$ need 
not have any physical significance.

Finally, for a complex chaotic attractor in general, the redistribution of the measures as it 
evolves in time 
can take place in more than two scales. Thus it appears that a characterisation based 
on only two scales is rather approximate as we 
tend to lose some information regarding the other scales involved.
But we have found that the $D_q$ and $f(\alpha)$ curves 
of a multi scale Cantor set can be mapped  onto that of an equivalent two scale 
Cantor set. 
These two scales 
may be functions of the actual scales involved and  may contain the missing information 
in an implicit way. 
Thus, an important outcome of the present analysis is the realisation that the 
dynamical information that can be retrieved from the $f(\alpha)$ spectrum is limited to 
only two scales. 
In this sense, a two scale Cantor measure can be considered as a 
good approximation to describe the multifractal properties of natural systems.

\begin{acknowledgments}
KPH  acknowledges the hospitality and computing facilities in IUCAA, Pune.
\end{acknowledgments}

\end{document}